\documentclass[pdflatex,sn-nature]{sn-jnl}% Math and Physical Sciences Numbered Reference Style 

\usepackage{graphicx}%
\usepackage{multirow}%
\usepackage{amsmath,amssymb,amsfonts}%
\usepackage{fdsymbol}
\usepackage{amsthm}%
\usepackage{mathrsfs}%
\usepackage[title]{appendix}%
\usepackage{xcolor}%
\usepackage{textcomp}%
\usepackage{manyfoot}%
\usepackage{booktabs}%
\usepackage{algorithm}%
\usepackage{algorithmicx}%
\usepackage{algpseudocode}%
\usepackage{listings}%
\begin{document}

\title{Competing incommensurability, electronic correlations, and superconductivity in a hybrid transition metal dichalcogenide}

\author*[1]{\fnm{Jean C.} \sur{Souza}}\email{Jean.Souza@weizmann.ac.il}
\equalcont{These authors contributed equally to this work.}

\author[1]{\fnm{Moshe} \sur{Haim}}
\equalcont{These authors contributed equally to this work.}

\author[2]{\fnm{Lorenzo} \sur{Crippa}}
\equalcont{These authors contributed equally to this work.}

\author[1]{\fnm{Hyeonhu} \sur{Bae}}
\equalcont{These authors contributed equally to this work.}

\author[1]{\fnm{Edanel} \sur{Fishbein}}

\author[4]{\fnm{Jonathan} \sur{Ruhman}}

\author[1,3]{\fnm{Binghai} \sur{Yan}}

\author[5]{\fnm{Amit} \sur{Kanigel}}

\author[6]{\fnm{Roser} \sur{Valentí}}

\author[1]{\fnm{Nurit} \sur{Avraham}}

\author*[1]{\fnm{Haim} \sur{Beidenkopf}}\email{Haim.Beidenkopf@weizmann.ac.il}

\affil[1]{\orgdiv{Department of Condensed Matter Physics}, \orgname{Weizmann Institute of Science}, \orgaddress{\city{Rehovot}, \postcode{7610001}, \country{Israel}}}

\affil[2]{\orgdiv{Institute of Theoretical Physics}, \orgname{University of Hamburg}, \orgaddress{\city{Hamburg}, \postcode{22607}, \country{Germany}}}

\affil[3]{\orgdiv{Department of Physics}, \orgname{The Pennsylvania State University}, \orgaddress{\city{University Park}, \state{PA}, \postcode{16802}, \country{USA}}}

\affil[4]{\orgdiv{Department of Physics}, \orgname{Bar Ilan University}, \orgaddress{\city{Ramat Gan}, \postcode{5290002}, \country{Israel}}}

\affil[5]{\orgdiv{Department of Physics}, \orgname{Technion - Israel Institute of Technology}, \orgaddress{\city{Haifa}, \postcode{32000}, \country{Israel}}}

\affil[6]{\orgdiv{Institut für Theoretische Physik}, \orgname{Goethe Universität Frankfurt}, \orgaddress{\city{Frankfurt}, \postcode{60438}, \country{Germany}}}

\abstract{
The engineering of superlattices in two-dimensional van der Waals materials has enabled the realization of rich phase diagrams hosting topological and strongly correlated phases. While incommensurability is widespread in three-dimensional systems, the role of moiré potentials in bulk materials remains largely unexplored. 
Here, using scanning tunneling microscopy, we demonstrate that a bulk transition-metal dichalcogenide polytype, 4Hb-TaS$_2$, hosts an emergent incommensurate potential between its alternating 1T and 1H layers. Interplay with a concomitant incommensurate charge-density wave suppresses the long-range order of this potential, leading to intricate coupling with electronic correlations in the doped 1T surface layer. 
Combining density functional theory with dynamical mean-field theory, we show that the lattice mismatch locally modulates the interlayer distance, thereby tuning both hybridization and charge transfer between the correlated 1T and metallic 1H layers. This redistribution of charge drives the system towards a doped Mott regime, in which the remaining local moments become self-screened, giving rise to a zero-bias resonance.
 We further find that bulk superconductivity competes with both the underlying landscape and the associated charge transfer. Our results establish incommensurate potentials as a previously overlooked ingredient in hybrid transition-metal dichalcogenides, highlighting their central role in the interplay between electronic correlations, charge-density-wave order, and unconventional superconductivity.
}

\maketitle

Incommensurability arises when a modulation cannot be expressed as a linear combination of the fundamental lattice wavelengths. Such behavior commonly appears in complex ordered phases, including charge-density waves (CDWs) and magnetic structures \cite{mcmillan1976theory,cheong1991incommensurate,miao2019formation}, and represents a fundamental manifestation of aperiodicity in quantum materials \cite{janssen2018aperiodic}. Because translational symmetry is lost, Bloch’s theorem no longer strictly applies, and the conventional Brillouin-zone description breaks down. This challenge can be resolved through the concept of superspace crystals, in which the aperiodic structure is viewed as the projection of a periodic lattice in a higher-dimensional reciprocal space \cite{de1974pseudo,janner1977symmetry,janner1980symmetrya,janner1980symmetryb,ozawa2019topological}.

Aperiodic lattices, including quasicrystals and misfit compounds, have been investigated for decades \cite{wiegers1996misfit,ng2022misfit,steinhardt1987physics,levine1986quasicrystals,bindi2009natural}. Recently, however, incommensurability has emerged as a powerful design principle in two-dimensional (2D) materials, where the stacking or twisting of atomically thin layers generates long-wavelength moiré potentials \cite{andrei2021marvels,mak2022semiconductor,nuckolls2024microscopic}. These superlattices reshape the electronic structure by folding the Brillouin zone and strongly renormalizing the bandwidth, thereby enabling correlated and topological phases ranging from Mott states to unconventional superconductivity. Although related aperiodic structures have recently been identified in composite bulk crystals \cite{devarakonda2024evidence,kurumaji2025electronic,nuckolls2025higher}, it remains largely unexplored whether moiré potentials can similarly control electronic correlations in three-dimensional materials.

The transition-metal dichalcogenide polytype 4Hb-TaS$_2$ provides a natural platform to address this question. This compound crystallizes in the hexagonal space group P6$_3$/mmc and consists of alternating 1T and 1H TaS$_2$ layers (Fig. \ref{Figtopos}a). In isolation, the 1T polymorph hosts a half-filled correlated state that forms a Mott insulator in the monolayer limit \cite{wang2020band,liu2021monolayer,aishwarya2022long,zhang2024quantum,huang2025doped,ruan2021evidence,butler2025topical}, whereas the 1H polymorph is a superconducting metal \cite{navarro2016enhanced,yang2018enhanced}. The intricate coupling between these subsystems results in a myriad of exotic states and ensuing phenomena, including its unconventional superconductivity and magnetic memory effects \cite{nayak2021evidence,persky2022magnetic,silber2024two,almoalem2024observation,yang2024signature}. Yet the electronic structure of this hybrid system remains poorly understood. Strong charge transfer (CT) from the 1T to the 1H layers appears to suppress the nominal half-filled Mott state, while additional states near the Fermi level cannot be captured within a simple single-particle picture \cite{almoalem2024charge}.

Here, we show that an emergent potential provides this missing ingredient. The pattern originates from the $\sim$ 1\% lattice mismatch between the 1T and 1H layers that modulates the electronic landscape in the bulk crystal. Using scanning tunnelling microscopy (STM) and spectroscopy (STS) on the 1T-terminated surface, together with density functional theory (DFT) and dynamical mean-field theory (DMFT), we demonstrate that a lattice mismatch locally tunes charge transfer and electronic correlations. In contrast to 2D moiré materials, where Brillouin-zone folding dominates, the incommensurate potential in 4Hb-TaS$_2$ primarily modulates interlayer hybridization and charge transfer between the 1T and 1H layers, thereby shifting the correlated flat band in the 1T layer towards the Fermi level. This local tuning produces an electronically screened state characterized by a zero-bias conductance peak accompanied by Hubbard-band satellites, consistent with a screening spectral signature \cite{vavno2021artificial,ayani2024probing,kumar2023first}. We further find that the previously reported abrupt charge transfer through a first-order quantum phase transition is induced by a realignment of the incommensurate potential, which is strikingly  interlinked with the superconducting transition, revealing a three-way interplay among CDW order, superconductivity, and the emergent lattice-mismatch landscape. These results establish moiré-CDW potentials as a previously overlooked tuning parameter in hybrid transition-metal dichalcogenides and extend the concept of incommensurate engineering beyond two-dimensional heterostructures to correlated bulk quantum materials.

\subsection*{Signatures of an underlying incommensurability}\label{moiré}

\begin{figure}[!ht]
\includegraphics[width=0.98\textwidth]{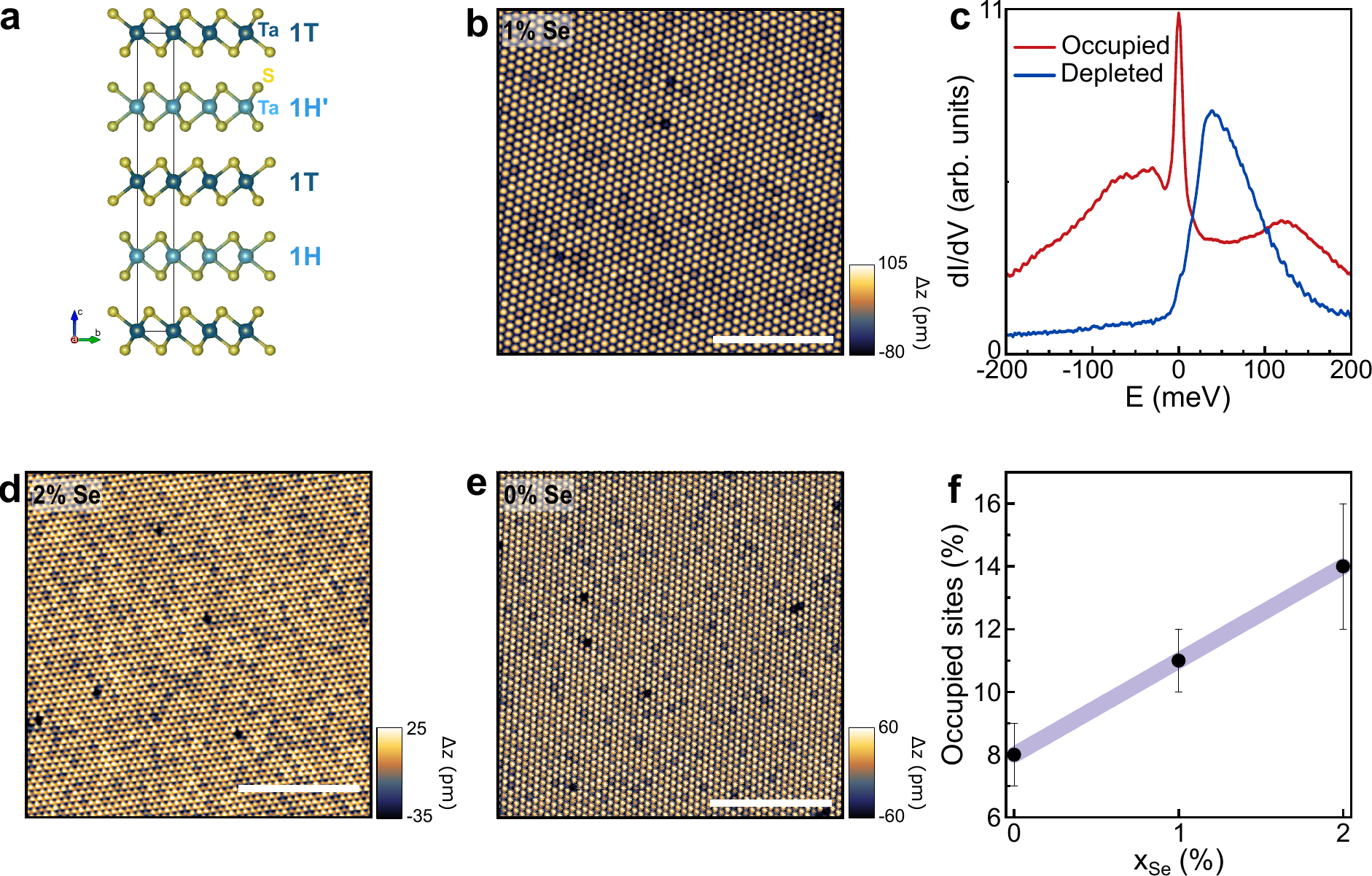}
\centering
\caption{\textbf{Intrinsic origin for undepleted CDW sites in 4Hb-TaS$_2$.} \textbf{a} Crystal structure of 4Hb-TaS$_2$. \textbf{b} Representative topography taken at $T$ = 4.2 K for $x_{Se}$ = 1\%. \textbf{c} Representative $dI/dV$ spectra of occupied and depleted CDW sites (red and blue lines, respectively) taken at $T$ = 4.2 K for $x_{Se}$ = 1\% ($V_{bias}$ =  200 mV, $I_{sp}$ = 200 pA, $V_{AC}$ = 5 meV). \textbf{d, e} Representative topographies for $x_{Se}$ = 2\%, 0\%. All topographies were taken with $V_{bias}$ = 100 mV and $I_{sp}$ = 50 pA, and the white scale bar is 20 nm. \textbf{f} Percentage of the total sites that are occupied as a function of Se doping, $x_{Se}$. The statistical analysis is further described in the SM. The solid purple line is a linear fit.} \label{Figtopos}
\end{figure}

Previous works speculated that Se substituted-defects were the origin of charge-transfer inhibition \cite{kumar2023first,yang2025nanoscale}. Introducing Se into the 4Hb-TaS$_2$ lattice also increases the interlayer 1H/1T distance ($c$/$a$ ratio) \cite{liu2014coexistence,xie2023revisit}, making it a relevant and interesting tuning parameter. To separate the role of Se versus that of the underlying lattice mismatch potential, we examined the Se substitution dependence of 4Hb-TaS$_{2-x}$Se$_x$ at the low- to no-doping limit. For that, we synthesized single crystals of 4Hb-TaS$_{2-x}$Se$_x$ with $x$ = 0\%, 1\%, and 2\% and cleaved them along the (001) plane. In topography, the 1T terminations in all cases exhibit the well-established $\sqrt{13}$ x $\sqrt{13}$ CDW (Fig. \ref{Figtopos}b) \cite{butler2020mottness,kumar2023first,butler2025topical}. The differential conductance spectrum (d$I$/d$V$), which is proportional to the density of states (DOS), shows at the majority of CDW sites the typical behavior of a depleted Mott state - a broad peak above the Fermi energy ($E$ $>$ 0) demonstrated in Fig. \ref{Figtopos}c (blue line) \cite{kumar2023first}. The charge depletion of the remote lower Hubbard band (LHB) is facilitated by the charge transfer from the 1T layer to the 1H metal, with an average, estimated by photoemission, to be $CT$ = 92 $\pm$ 11\% of electrons per CDW site in the bulk \cite{almoalem2024charge}. However, atomic-scale STS finds that a small concentration of 11 $\pm$ 1\% of the CDW sites hosts a different spectrum featuring well-defined filled LHB below the Fermi energy along with an empty upper Hubbard bands (UHB) above, and a zero-bias conductance peak (ZBCP) \cite{kumar2023first}. The distinct spectroscopic features are also reflected in the topography, as the integrated DOS from the Fermi energy up to the parking bias is modified by the shift of the LHB from below to above the Fermi energy as it becomes depleted. In the depleted Mott case, the CDW sites appear to have a higher topographic height than the resonance spectrum of the occupied CDW sites, as demonstrated in Figs. \ref{Figtopos}b, \ref{Figtopos}d and \ref{Figtopos}e. Therefore, these CDW sites are also known as bright and dim sites, respectively. 

As already mentioned, previously, the undepleted spectrum obtained on some of the CDW sites of the 1T-terminated surface of 4Hb-TaS$_{2-x}$Se$_x$ has been commonly attributed to the presence of $x =$ 1\% Se atoms in the single crystal growth, which was thought to inhibit charge transfer \cite{kumar2023first,yang2025nanoscale}. Indeed, about $x \approx$ 1\% Se concentration was seen to dramatically stabilize the growth and improve sample quality. 
In line with the Se-defect reasoning, varying the Se concentration appears to proportionally affect the concentration of undepleted sites. We thus measured several independent surfaces of samples with varying Se concentration (Supplementary Note S.1) and obtained that their occupied sites concentration increased linearly, as $n_{occ}(x) \approx 0.08 + 3x$, with Se-concentration, $x$, shown in Fig. \ref{Figtopos}f. Strikingly, however, the non-zero intercept implies undepleted CDW sites were still present for samples without any Se (Fig. \ref{Figtopos}e).

\begin{figure}[!ht]
\includegraphics[width=0.99\textwidth]{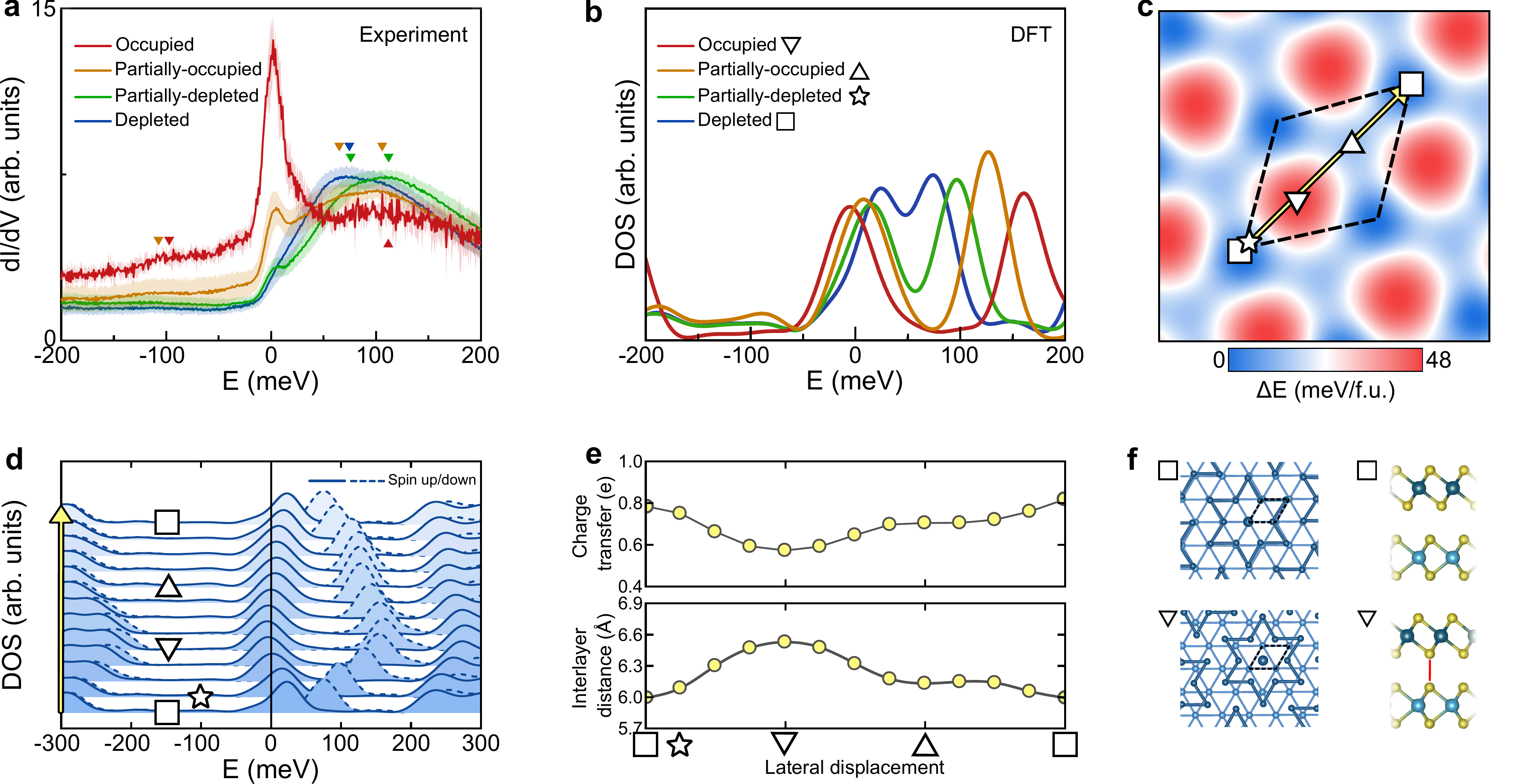}
\centering
\caption{\textbf{Modulation of the charge transfer and the hybridization strength by the lattice mismatch.} \textbf{a} Clustered averages of four different types of spectra obtained at the same 1T surface ($V_{bias}$ =  120 mV, $I_{sp}$ = 150 pA, $V_{AC}$ = 5 mV). The shaded margins represent the standard deviation for each spectrum. \textbf{b} DFT calculated DOS for the 1T layer of bilayer 1T/1H-TaS$_2$ in selected lateral displacements (marked in \textbf{c}). The labels denote the experimental spectral archetypes; in the DFT calculations, they identify representative local registries with different residual fillings of the 1T-derived flat band. \textbf{c} Relative energy landscape $\Delta E$ for lateral displacement of a 1T layer on a fixed 1H layer. Energies are referenced to the lowest energy registry. The black dashed rhombus denotes the moiré primitive cell. The square represents the stable configuration, while both triangles and the star are metastable ones of higher symmetry. \textbf{d} spin-polarized DOS on the 1T termination along the arrow in \textbf{c}. \textbf{e} Relationship between the interlayer distance and the charge transfer in the 1T/1H bilayer system. \textbf{f} Top and side views of the 1T/1H bilayer structures with lateral displacements of $\sqrt{13}$ x $\sqrt{13}$ for further CDW order of the 1T layer with respect to the 1H layer. Ta atoms in 1T and 1H layers are depicted as dark and light blue spheres, respectively. In the top views, S atoms are omitted for clarity. Each case shows minimum and maximum interlayer distances, corresponding to the maximum and minimum charge transfers, respectively.}
\label{FigDFT}
\end{figure}

This observation points to an intrinsic mechanism, independent of Se substitution, that locally inhibits charge transfer and induces occupied screened CDW sites in the 1T layer of 4Hb-TaS$_2$. Given the natural lattice mismatch between the 1T and 1H layers, an incommensurate potential provides a plausible origin for this intrinsic modulation as charge transfer may very well be affected by the inter-alignment of the S orbitals. The mismatch between lattice constants of bulk 1T- and 2H-TaS$2$, reported as 3.38 \AA\ and 3.342 \AA, respectively \cite{bin2007anisotropic}, is of 1.1\% mismatch. This corresponds to a moiré unit cell that spans about 550 CDW sites, a portion of which will have S stacking less favorable for charge transfer. The intrinsic density of undepleted sites in Se-free samples finds that within each moiré cell about 8\% of sites will misaligned sufficiently to inhibit charge transfer.
While moiré superstructures commonly produce ordered or quasi-ordered superlattices, with periodic electronic localization and band folding \cite{uri2023superconductivity,guo2025synthesis}, the incommensurate potential in our system does not exhibit long-range order, as verified by the Fourier transform of the topography (Supplementary Note S.2). We attribute the complex structure of the underlying landscape to its competition with the strong incommensurate $\sqrt{13}$ x $\sqrt{13}$ CDW order, which, together with local relaxation, results in an irregular pattern. 

An accurate spectroscopic inspection further reveals a slightly more intricate scheme beyond the well-appreciated binary view of depleted and occupied (hosting ZBCP) archetypes of CDW sites (Fig. \ref{FigDFT}a). Instead of continuous evolution, we identify a discrete set that includes exactly two additional intermediate configurations between depleted and occupied spectra. These two intermediate spectra show the development of the ZBCP, and concomitantly, a slightly more pronounced LHB (Fig. \ref{FigDFT}a). We identify them as partially depleted and partially occupied based on their relative LHB-to-UHB occupation and the relative intensity of the ZBCP they exhibit. We also note that the upper remote band appears to comprise two adjacent kinks (marked by arrows). Their coexistence and merger with STM tip height was previously noted in Fig. 4b (dotted lines) of Ref.\cite{kumar2023first}. 

To further explore the intricate interplay between the electronic band structure and the lattice mismatch, we performed density functional theory (DFT) calculations on a 1T/1H bilayer. To approximate the incommensurate 1T/1H mismatch, we sampled local registries by laterally displacing the 1T layer relative to the fixed 1H layer and relaxing the ionic coordinates. This local-registry approach does not simulate a full incommensurate moiré supercell; rather, it extracts the local charge-transfer and flat-band-filling trends expected within such a landscape.
Based on our relative energy calculations presented in Fig.\ref{FigDFT}c, one stable (square symbol), two metastable (pyramid and inverted pyramid), and one intermediate position (star) across the moiré unit cell (dashed lines) were identified. The relative energy difference between the stable configuration (square) and the metastable ones is $\Delta_{E}$ = 11, 11, and 48 meV per formula unit for the star, pyramid, and inverted pyramid positions, respectively. The calculated local 1T DOS at these positions unveils stacking-dependent flat band energies and occupations, reflecting the modulation of charge transfer from the 1T to the 1H layer (Fig.\ref{FigDFT}b and Supplementary Note S.7). 
We relate these four stable and metastable positions to the four spectra archetypes obtained experimentally (Fig. \ref{FigDFT}a). The near-$E_\mathrm{F}$ spectral weight in the spin-polarized DFT DOS originates from partially occupied, statically exchange-split 1T-derived flat band that differs from the ZBCP observed in STS, which is a dynamical many-body feature of the finite-doped correlated orbital.

A deeper understanding can be obtained by examining the spatially extracted DFT linecuts connecting stable and metastable positions of the moiré unit cell, as exemplified in Fig. \ref{FigDFT}d. Here, we observe a modulation of the flat band position and, consequently, of the DOS at the Fermi level. As the 1T layer is displaced with respect to the 1H one, there is an increase in the filling factor and, hence, an enhancement of the spin splitting (solid versus dotted lines). It is worth noting that the largest DOS at the Fermi level is obtained in the inverted pyramid metastable position. As such, we associated the stable and metastable positions with flat-band filling, specifically with the depleted (square), partially depleted (star), partially occupied (pyramid), and occupied (inverted pyramid) spectra.

As demonstrated in Fig. \ref{FigDFT}e, the 1T/1H lateral displacement (Fig. \ref{FigDFT}f) indeed reduces charge transfer due to a larger interlayer distance. The calculated most suppressed charge transfer is $CT$ $\sim$ 0.6 $e$ for the occupied configuration (Fig. \ref{FigDFT}e). It is larger, yet comparable to the estimates for the 1T/1H bilayer samples \cite{crippa2024heavy}. Our DFT calculations thus indicate that the charge-transfer modulation can be interpreted as an increase in the interlayer distance due to incommensurability (Fig. \ref{FigDFT}e). One contributing factor is the reduced overlap between the $d$ orbitals of the 1T and 1H layers as their separation increases, which effectively suppresses charge transfer (Fig. \ref{FigDFT}f). It is worth noting that recent DFT results show that, among 90 possible defects, only a few highly energetic ones can alter charge transfer \cite{karbasizadeh2026revealing}. As such, this incommensurate potential provides a natural explanation for the experimentally observed modulation.
 Although the single particle picture captures the intricate interplay between the incommensurate lattice and electronic band structure, the calculated density of states does not fully reproduce the experimentally measured d$I$/d$V$ spectra (Figs. \ref{FigDFT}b and \ref{FigDFT}a), implying a significant role for many-body interactions.

\subsection*{Interactions and the zero-bias conductance peak}\label{Mott}

\begin{figure}[!ht]
\includegraphics[width=0.99\textwidth]{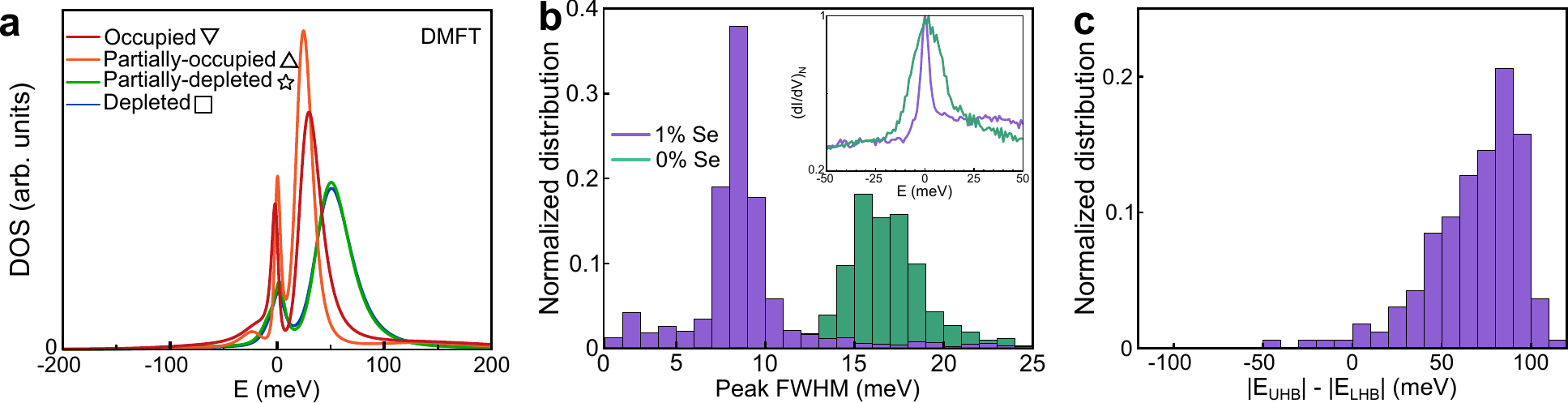}
\centering
\caption{\textbf{The origin of the screening metal. a} DMFT calculated spectra on the 4 (meta) stable moiré positions identified in DFT (marked by corresponding symbols in Fig. \ref{FigDFT}c). \textbf{b} Comparison between the ZBCP width for $x_{Se}$ = 0\% and 1\%. Around 400 occupied sites spectra  were analyzed for $x_{Se}$ = 0\% and about 600 sites for $x_{Se}$ = 1\%. The inset shows representative ZBCP spectra from both sample types. \textbf{c} Asymmetry of the LHB and the UHB energy relative to $E_{F}$ for $x_{Se}$ = 1\%. About 250 occupied spectra were analyzed. \label{Figorigin}}
\end{figure}

The partial charge transfer from the 1T layer to the 1H layer places the 1T-derived flat band in a finite hole-doped (Mott) regime, providing an intriguing platform for studying correlated flat band physics \cite{ruan2021evidence,ayani2024probing,ayani2024unveiling,wan2023evidence,shen2022inducing,vavno2021artificial,crippa2024heavy,huang2025doped,bae2025designing}. The ZBCP, accompanied by Hubbard-band satellites, clearly indicates the presence of strong correlations. Two possible mechanisms behind these features are either charge-carrier depletion in the 1T layer through the (partial) charge transfer from 1T to 1H, or screening of the local moments in the 1T layer by the metallic electrons of the 1H layer~\cite{crippa2024heavy}. In the first case, the ZBCP is due to the presence of heavily renormalized quasiparticle excitations in the partially doped 1T layer. In the second, the ZBCP is due to Kondo screening of the local 1T moments by the 1H metal.
Although previous results indicate that the second effect is the dominant one in 1T/1H-TaS$_2$ bilayers, 
it remains unclear whether the same holds for bulk 4Hb-TaS$_2$. For that, we need to elucidate how the lattice-mismatch landscape controls the ZBCP width and intensity, as well as the positions and asymmetries of the Hubbard-band satellites, across the four archetype spectra.
Quantifying the relative contribution of each mechanism is essential for identifying the local correlated state.

To properly assess this, we have conducted a DMFT study based on the DFT simulations: for each of the representative local registries in Fig.~\ref{FigDFT}c, we obtained a single-particle DFT-based tight-binding Hamiltonian (Methods) describing a Periodic Anderson model~\cite{crippa2024heavy}. In this model, the physics of the narrow and dispersive bands is encoded in two different flavors of electrons: correlated ($d$) and uncorrelated ($c$). The DFT-derived charge transfer sets the residual filling of the correlated 1T-derived orbital. Likewise, the effect of interlayer coupling is modeled via a $d-c$ hybridization term, depending on a scalar coefficient $V_{0}$, which keeps the itinerant 1H states explicitly coupled to it. Hence, the parameter $V_0$, which is related to the stacking-dependent effective amplitude, is a crucial control knob that determines the properties of the spectrum.

In Fig. \ref{Figorigin}a we show the local spectral function resulting from our DMFT simulations. The most striking feature is the presence of a persistent narrow resonance at zero energy bias. The DMFT peak appears on a much smaller energy scale than the spectral maxima of DFT simulations (Fig.~\ref{FigDFT}b).
As detailed in Table~S.6, all configurations feature a considerable charge transfer between the 1T and 1H layers.%, which allows us to conclude for the quasiparticle origin of the resonance.
It is instructive to examine the evolution of the ZBCP with Se doping through its effect on the model parameters. Incorporating Se increases the lattice constant along the $c$ axis of 4Hb-TaS$_2$, an effect which we visualize directly through crystallographic step-edges between 1T and 1H surface terminations (Supplementary Note S.3) \cite{liu2014coexistence,xie2023revisit}. 
In turn, bringing the 1T and 1H layers apart affects both the hybridization between 1T and 1H and charge transfer. Indeed, when we histogram the experimental ZBCP width, shown in Fig. \ref{Figorigin}b, we systematically obtain narrower ZBCP for $x_{Se}$ = 1\% than in 0\% samples (demonstrated by the inset). 
As for its intensity, DMFT calculations find that the ZBCP is enhanced with decreasing charge transfer, as seen in Fig.~\ref{Figorigin}a where the resonance in the depleted configuration ($CT=0.79$) is remarkably more smeared than in the occupied configuration ($CT=0.57$). However, beyond verifying the consistency between the experimental observation and DMFT modeling, both, increased hybridization and increased charge doping will potentially broaden the ZBCP. Therefore, the evolution of the ZBCP with Se doping alone cannot resolve its exact origin.

We thus turn to the evolution of the Hubbard bands. By effectively modifying the layer separation in the DMFT simulations, as in Fig. \ref{Figorigin}a, we observed that the reduction of charge transfer at the occupied sites is accompanied by a slight shift of the UHB away from the Fermi level when compared to the partially-occupied configuration. This is even more evident in the depleted cases, in agreement with the observed trend in the experimental spectra. 
By analyzing over 250 occupied sites spectra within a large field of view of a $x_{Se}$ = 1\% sample, we find a clear and significant asymmetry of about 80 meV in the energy barycenter of the UHB and LHB, as shown in Fig. \ref{Figorigin}c. Upon increasing charge transfer from the 1T to the 1H layer, the upper Hubbard band splits further away from the ZBCP, while the lower one shifts only slightly away from the Fermi level. It is also immediately noticeable how the spreading and overall spectral weight substantially differ between the UHB and the LHB.
 
By contrast, in the absence of charge transfer, increasing hybridization does not generate asymmetry: in this case, both Hubbard bands have comparable spectral weight and are positioned symmetrically around the Fermi level \cite{crippa2024heavy}. 
We thus conclude that the asymmetric spectral distribution of the remote bands originates in self-screening by charge doping of the 1T Mott state, rather than hybridization with the 1H metal. 

\subsection*{Interplay between incommensurability and superconductivity}\label{quantum}

\begin{figure}[!ht]
\includegraphics[width=0.99\textwidth]{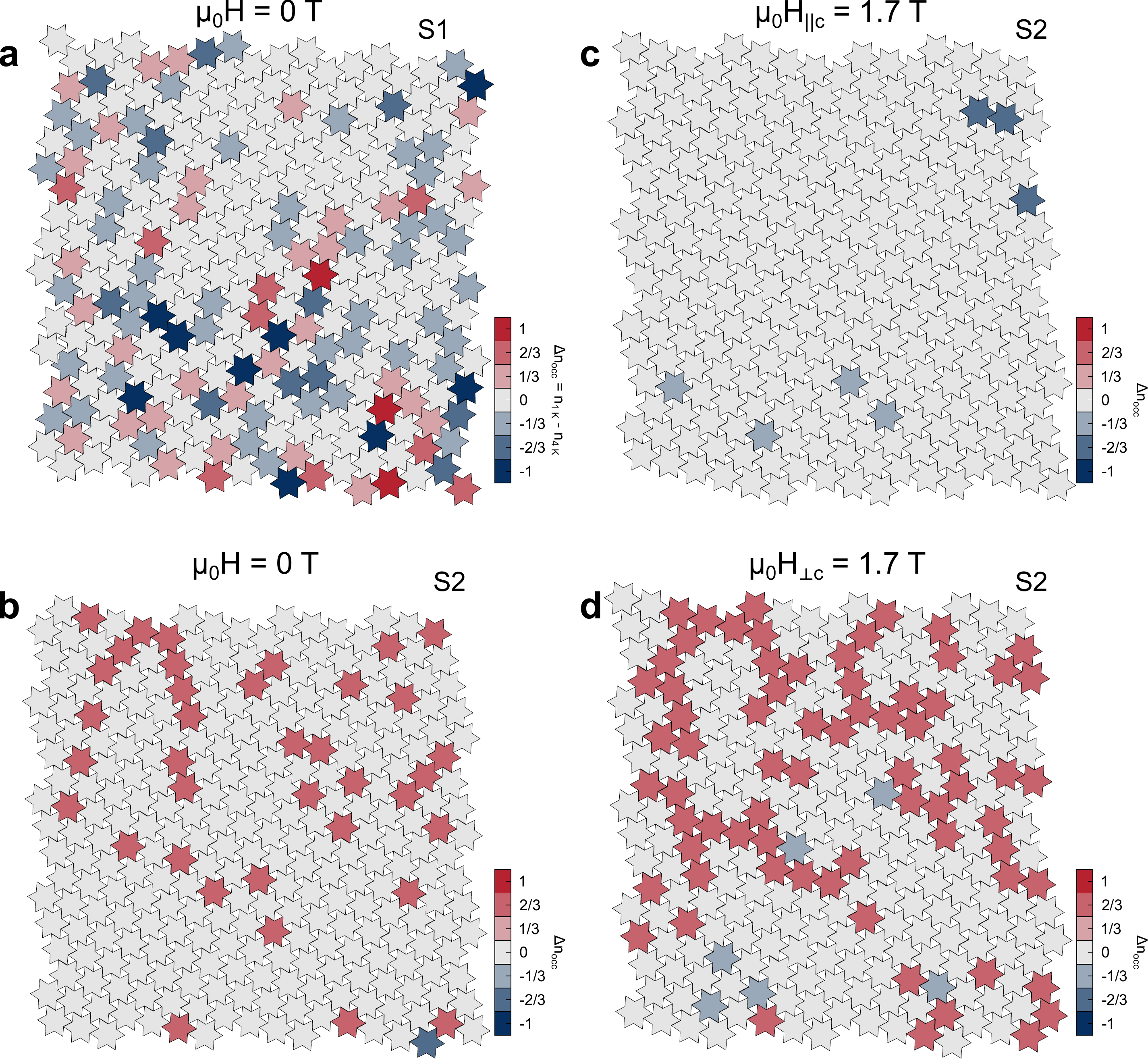}
\centering
\caption{\textbf{The effect of superconductivity on the breathing transition of the incommensurate potential.} \textbf{a} Change of the enumerated `occupation level' of the CDW sites $\Delta n_{occ}$ between $T$ = 4.2 K to $T$ = 1.1 K for sample S1. Change of the `occupation level', $\Delta n_{occ}$, for \textbf{b} $\mu_{0}H$ = 0 T, \textbf{c} $\mu_{0}H_{\parallel c}$ = 1.7 T and \textbf{d} $\mu_{0}H_{\perp c}$ = 1.7 T in sample S2. The measurements were first taken at $T$ = 4.2 K with the respective applied magnetic fields, then the sample was cooled down with the applied field for the measurements at $T$ = 1.1 K. All samples had $x_{Se}$ = 0.} \label{Figchanges}
\end{figure}

After establishing the role of the lattice-mismatch landscape in charge transfer, interlayer hybridization, and the many-body origin of the spectra, we turn to its interplay with superconductivity. Previously, it has been established that there is a first-order quantum phase transition between occupied and depleted CDW configurations \cite{kumar2023first}. This phase transition is induced either by changing the temperature or the tip-sample distance \cite{kumar2023first,yang2025nanoscale}. Intriguingly, the first-order phase transition occurs at temperatures in the vicinity of the superconducting transition, $T_{C}$ $\approx$ 2.6 K \cite{kumar2023first}. However, broader statistics on the commonality of occurrence of the transition were lacking in previous studies. Here, we provide further insight by gathering the statistics on spectral site changes spatially for $x_{Se}$ = 0. We obtain spectroscopic d$I$/d$V$ mappings at the same field of view at temperatures of 4.2 K and 1.1 K and the effect of applying a magnetic field on them, given in Fig. \ref{Figchanges} (Supplementary Note S.4). 

We first examined the wave function distribution across the CDW sites. By mapping the zero-bias DOS, we easily distinguish occupied sites among the majority of depleted ones at 4.2 K and 1.1 K (Supplementary Note S.4). Partially occupied or partially depleted sites are more challenging to distinguish. For accurate identification, we clustered the STS of each point in the d$I$/d$V$ maps into one of five groups: four for the different archetypes and one for impurities (Supplementary Note S.4). For a more detailed quantification of the phase transition, we enumerate the `occupation level', $n_T$, of each CDW site, where $n_T$ = 0, 1/3, 2/3, and 1 correspond to depleted, partially depleted, partially occupied, and occupied sites, respectively. Figure \ref{Figchanges}a highlights the overall change of the `occupation level' $\Delta n_{occ}$ = $n_{1.1 K}$ - $n_{4 K}$ for sample S1. About 40\% of the 280 CDW sites changed their archetype when we changed the temperature. While $\sim$ 75\% of the partially filled states change, depleted and occupied sites were more stable, with only $\sim$ 25\% changing their occupation. Furthermore, whereas only 3 sites had a full change in their `occupation level' ($\Delta n_{occ}$ = $\pm$ 1), the vast majority (68\%) of changes involved only $\Delta n_{occ}$ = $\pm$ 1/3. This is consistent with the interpretation of the spectra being intricately linked to stable and metastable positions. Although there is a net increase of about 15\% in the summed `occupation level', it is important to note that locally there are changes in both ways. The changes are not equally distributed across the field of view, as they seem to accumulate along particular domains. Yet, the bunched changes in occupation levels are commonly of opposite trends (negative or positive). Therefore, at first glance, the change in occupation level as a function of temperature could be related to breathing of the lattice mismatch, driven by the entropy release of a phase transition.

While the changes between 4.2 K and 1.1 K are abundant, it remains unclear whether the competing energy scale is the superconducting transition at $T_c \approx 2.6$ K. To explore this further, we measure spectroscopic d$I$/d$V$ maps for another sample (sample S2) at $T$ = 4.2 K and 1.1 K (Supplementary Note S.4). Contrary to the first sample, the summed `occupation level' of this surface decreases at base temperature at zero applied magnetic field (Fig. \ref{Figchanges}b), making it unlikely that any origin of the changes is related to a temperature-induced electronic screening transition. Strikingly, the changes in occupation are heavily suppressed when cooled below $T_c$ in the presence of an out-of-plane magnetic field $\mu_0H$ = 1.7 T (Fig. \ref{Figchanges}c), which is above the critical magnetic field of $\mu_0H_{c2}$ $\sim$ 1.1 T (Supplementary Note S.5). When repeated with an in-plane field of the same magnitude (Fig. \ref{Figchanges}d), which is below the in-plane critical field of $\sim$ 12 T, changes in occupation are recovered. Again, the changes in occupation level accumulate along well-defined directions, in agreement with a change in the incommensurate potential. Our results thus indicate a significant interplay between the superconducting transition and the lattice mismatch at the first-order quantum phase transition.

\subsection*{Discussion and Conclusion}\label{secdisc}

Recently, there have been reports of other Ta-based TMDs misfits and polytypes \cite{devarakonda2024evidence,nuckolls2025higher,almoalem2025mixed}, that may have been affected by the lattice mismatch. However, the correlated nature of the insulating 1T layer in the 4Hb-TaS$_2$ seems to signal the effect of the lattice mismatch that locally tunes and modulates the strength of electronic correlations. Beyond demonstrating the interplay between electronic correlations and incommensurability in a bulk crystal, our results also provide insights into previous 1T/1H bilayer systems \cite{ruan2021evidence,ayani2024probing,ayani2024unveiling,wan2023evidence,shen2022inducing,vavno2021artificial,huang2025doped}. Our DMFT results show that the ZBCP persists even for significant charge-transfer values (Fig. \ref{Figorigin}a). While the ZBCP may develop into a Fano line-shape through co-tunneling between the conduction electrons and localized states\cite{madhavan1998tunneling}, the remote LHB and UHB are generally not affected by this process \cite{vzitko2007many,huang2021tunneling,huang2023universal,van2024modulated,bagchi2024spin}. Therefore, we focused mainly on the position of the satellite peaks. We demonstrate that their asymmetry arises whenever the 1T layers are doped away from half-filling, with each CDW site, on average, hosting less than one electron \cite{janivs2008kondo}. We stress that almost all the 1T/1H heterostructures reported so far show a strong asymmetry of the LHB and UHB relative to the Fermi level \cite{ruan2021evidence,ayani2024probing,ayani2024unveiling,wan2023evidence,shen2022inducing,vavno2021artificial,huang2025doped} in agreement with our observation here. Therefore, this should be regarded as an experimental hallmark of the inevitable 1T hole doping resulting from charge transfer to the 1H layer. In this scenario, a proper description of the system must account for the two species of charges (localized and delocalized) within the 1T layer, leading to a doped Anderson or a doped extended Hubbard model. Due to hopping between CDW sites induced by doping, we have two possible screening channels. The first one is the traditional Kondo, which relies on the hybridization between 1H conduction electrons and 1T localized moments, forming singlet quasiparticles. The second is the electronic screening between the 1T localized and delocalized states.

Another puzzling result arises from the interplay between the underlying potential and the superconducting state, which has a gap of $\sim$ 0.6 meV (Supplementary Note S.5). While the charge-transfer amplitude remains unchanged, the incommensurate landscape is modified, leading to the perceived first-order quantum phase transition \cite{kumar2023first,yang2025nanoscale}. At first glance, one might assume that the thermal expansion of the 1H layer relative to 1T across the superconducting transition could be the tipping point for the phase transition. However, the entropy release at the transition $T_{C}$ is about 30 mJ/mol K \cite{ribak2020chiral}, much lower than the DFT-calculated relative energy differences between configurations (a few tens of meV per formula unit). As such, the transition appears to be electronic, with the energy scale of the superconducting gap occasionally inducing a change between configurations. Another possible scenario could involve the spatial modulation of the superconducting gap function that would compete with the incommensurate landscape, leading to a reorganization of the archetypes \cite{guo2025synthesis}. Although modulations of the superconducting gap have not been detected yet in 4Hb-TaS$_2$, TMDs have shown pair density wave ground states \cite{agterberg2020physics,liu2021discovery,laskowski2025josephson}. Additionally, fingerprints of $q$ $\neq$ 0 pairing have been reported in 4Hb-TaS$_2$ \cite{nayak2021evidence,silber2024two,yang2024signature}. Given that the interplay between superconductivity and density waves remains an open question in many systems, ranging from UTe$_2$ to nickelates \cite{laskowski2025josephson,zhang2020intertwined,gu2023detection}, our results offer hints on how additional energy scales, such as the moiré potential, affect the intricate competition between density waves and superconductivity. Further Josephson spectroscopy would be particularly important for a better understanding of the origin of this transition.

\subsection*{Acknowledgments}

We acknowledge fruitful discussions with Beena Kalinsky and Giorgio Sangiovanni. We thank Tim Wehling for providing his tight-binding code. J.C.S. acknowledges support from the Paulo Pinheiro de Andrade fellowship. R.V. acknowledges support by the Deutsche Forschungsgemeinschaft (DFG,
German Research Foundation) for funding through projects TRR 288 — 422213477 (projects A05, B05) and through QUAST-FOR5249 - 449872909 (project P4).  L.C. acknowledges support from the Cluster of Excellence ‘CUI: Advanced Imaging of Matter' – EXC 2056 (Project No. 390715994), and SPP 2244 (WE 5342/5-1 project No. 422707584).

\subsection*{Authors contribution}
J.C.S. and M.H. acquired the STM data, with support from N.A. and H.B. J.C.S., M.H., and E.F. analyzed the STM data, with inputs from N.A. and H.B. H.B. and B.Y. performed the DFT calculations. L.C. and R.V. performed the DMFT calculations. A.K. provided the single crystals. J.C.S., M.H., L.C., R.V., and H.B. wrote the manuscript with substantial contributions from all authors.

\subsection*{Competing interests}
The authors declare no competing interests.

\bibliography{TaS2_refs}

\subsection*{Methods}\label{secmethods}

\bmhead{Sample growth}

\bmhead{STM measurements}

4Hb-TaS$_2$ single crystals were cleaved under ultrahigh-vacuum conditions at room temperature and immediately inserted into the STM head. We performed measurements for $x_{Se}$ = 0\%, 1\%, 2\% and 5\%. We cleaved 8 samples in total. The measurements were performed using Pt-Ir tips, which were characterized in a freshly prepared Cu(111) single crystal. This preparation ensured a stable tip with reproducible results. Topographies were collected in constant-current mode, with an initial current set point $I_{sp}$ and an applied bias voltage $V_{bias}$ to the sample. Spectroscopic measurements were obtained using standard lock-in techniques. Unless stated otherwise, we used a frequency $f$ = 793 Hz.

\bmhead{DFT calculations and tight-binding model}

Spin-polarized density functional theory (DFT) calculations were performed using the Vienna Ab initio Simulation Package (VASP) with the projector augmented-wave (PAW) method \cite{kresse1999ultrasoft}. We employed the Perdew-Burke-Ernzerhof (PBE) exchange-correlation functional \cite{perdew1996generalized}, and included van der Waals interactions using the zero-damping DFT-D3 scheme \cite{grimme2010consistent}. Electronic correlations of Ta $5d$ states in the 1T-TaS$_2$ and 1H-TaS$_2$ layers were treated within the Dudarev DFT+U approach \cite{dudarev1998electron}, using $U^\textrm{1T}_\textrm{Ta,5d}$=1.76 eV and $U^\textrm{1H}_\textrm{Ta,5d}$=2.82 eV, respectively \cite{ayani2024unveiling}. A plane wave kinetic energy cutoff of 400 eV was used. The Brillouin zone was sampled with an $8\times8$ $\Gamma$-centered k-point mesh, and Gaussian smearing with $\sigma=$30 meV was adopted for Brillouin zone integrations. The electronic energy convergence criterion was set to $10^{-7}$ eV. Structural relaxations were performed until the residual forces on each atom were less than 1 meV/\AA{}.

Starting from the band structures obtained from DFT, we have obtained a set of tight-binding models describing the low-energy physics of the system (see Fig. S.5). Similar to the one employed in ~\cite{crippa2024heavy}, these consist of three terms describing the correlated and uncorrelated electrons as well as their hybridization.
The Hamiltonian of the uncorrelated electrons has the form
\begin{equation}
    H_{1H} = \epsilon_{1H}\sum_{i\alpha}c^{\dagger}_{i\alpha}c_{i\alpha} + t_{H}\sum_{\langle i\alpha,j\beta\rangle} c^\dagger_{i\alpha}c_{j\beta}
\end{equation}
where $i,j$ represent nearest-neighbor sites and $\alpha,\beta$ run over 13 uncorrelated levels describing the $d_{z^2}$ orbitals of the $1H$ layer, which most contribute to the hybridization.
The correlated electrons are instead described by a single $d$ orbital following 
\begin{equation}
    H_{1T} = \epsilon_{1T}\sum_{i\alpha}d^{\dagger}_{i}d_{i} + t_{1T_{1}}\sum_{\langle i,j\rangle} d^\dagger_{i}d_{j} + t_{1T_{2}}\sum_{\langle\langle i,j\rangle\rangle} d^\dagger_{i}d_{j} + t_{1T_{3}}\sum_{\langle\langle\langle i,j\rangle\rangle\rangle} d^\dagger_{i}d_{j}
\end{equation}
where the hopping elements and sum go over first, second, and third nearest neighbors.
The hybridization, mediated by the 1H layer Ta atom identified by the index $\alpha=1$, has the form

\begin{equation}
H_V = V_0 \sum_i d_i^{\dagger} c_{i,1} + \text{h.c.}
\end{equation}
where $V_{0}$ is a scalar coefficient depending on the interlayer distance. We use the coefficients of $H=H_{1H}+H_{1T}+H_{V}$ as variational parameters to optimize a fit of the low-energy narrow bands derived by the DFT simulations. This gives access to a set of four tight-binding models which, together with the charge-transfer value also derived by DFT, form the starting point for the DMFT analysis. The resulting optimized parameters are shown in Table~S.5

\bmhead{DMFT calculations}
We performed Dynamical Mean-Field theory simulations over a periodic Anderson model made up of the previously defined tight-binding Hamiltonian plus a Hubbard interaction term
\begin{equation}
 H_U = U \sum_{i}n_{i\uparrow}n_{i\downarrow}
\end{equation}
where $U=100meV$ is the value of the local interaction, limited on the $d$-orbitals of the $1T$ layer. 
Numerical simulations were performed with the continuous-time Quantum Monte Carlo solver \textit{w2dynamics}~\cite{wallerberger2019}, at temperatures varying between $\sim230$ K and $\sim8$ K for each configuration. A maximum number of 25 self-consistency loops was found to yield convergence at all the chosen temperatures. The value of the charge transfer obtained via DFT simulation was enforced in the DMFT calculations by means of a self-consistently determined double-counting potential for the correlated orbital, adjusted loop-on-loop to ensure a $d$-orbital occupation of $1-CT$, where $CT=0$ signifies half-filling.

\subsection*{Data availability}
All data available can be obtained from the corresponding authors upon reasonable request.

\bmhead{Supplementary information}

The online version contains supplementary materials available at xxx.

\end{document}